# EOSFuzzer: Fuzzing EOSIO Smart Contracts for Vulnerability Detection[*][#]

Yuhe Huang
School of Computer Science and Engineering
Beihang University
Beijing, China
huangyuhe@buaa.edu.cn

Bo Jiang[†]
School of Computer Science and Engineering
Beihang University
Beijing, China
jiangbo@buaa.edu.cn

W.K. Chan
Department of Computer Science
City University of Hong Kong
Hong Kong
wkchan@cityu.edu.hk

## ABSTRACT

EOSIO is one typical public blockchain platform. It is scalable in terms of transaction speeds and has a growing ecosystem supporting smart contracts and decentralized applications. However, the vulnerabilities within the EOSIO smart contracts have led to serious attacks, which caused serious financial loss to its end users. In this work, we systematically analyzed three typical EOSIO smart contract vulnerabilities and their related attacks. Then we presented EOSFuzzer, a general black-box fuzzing framework to detect vulnerabilities within EOSIO smart contracts. In particular, EOSFuzzer proposed effective attacking scenarios and test oracles for EOSIO smart contract fuzzing. Our fuzzing experiment on 3963 EOSIO smart contracts shows that EOSFuzzer is both effective and efficient to detect EOSIO smart contract vulnerabilities with high accuracy.

## 1 Introduction

The blockchain technology is proposed as a value transfer network among peers with limited trust. Currently, the blockchain technology has evolved as a trust machine with which untrusted peers can cooperate with each other. Most of the current blockchain platforms supports smart contracts with which developers can build Decentralized Applications (DApps). Among them, the ecosystem of EOSIO platform is growing steadily. The types of DApps range from game, gambling, decentralized exchanges, tools, microblogging, etc.

However, the vulnerabilities within EOSIO smart contracts have led to non-trivial financial loss to its end users. For example, the random number generation vulnerabilities within the EOSIO gambling games (including EOSBet, EOSCast, FFGame, EOSDice, EOSWin, etc.) have led to the loss of around 170K EOS tokens [28]. The Fake EOS Transfer vulnerability within the EOSCast smart contract has led to the loss of around 60K EOS tokens [26]. The Forged Transfer Notification vulnerability within EOSBet has led to the loss of 140K EOS tokens [27]. Based on the price of the EOS tokens (about $5 on average) at the time of the attack, the accumulated amount of loss by these three vulnerabilities alone was around 1.9 million worth of USD. Therefore, building effective vulnerability detection tools for EOSIO smart contracts is valuable. However, no existing fuzzing tool is available to detect EOSIO smart contract vulnerabilities.

Different from the fuzzing of the Ethereum smart contracts, there are two main challenges with the fuzzing of EOSIO smart contracts. First, the EOSIO smart contracts adopt a unique EOS token transfer and notification mechanism through the *eosio.token* system smart contract, which can lead to distinctive EOSIO smart contracts vulnerabilities. To effectively trigger such vulnerabilities, we must design new attacker agents to trigger such vulnerabilities as well as new test oracles to precisely detect such vulnerabilities. Second, EOSIO smart contracts are executed by the WASM virtual machine, which supports different bytecode instruction sets from EVM. Furthermore, new types of vulnerabilities will continue to emerge with the development of EOSIO platform. Therefore, we must design the instrumentation scheme within the WASM VM in a general way to capture adequate information to support the detection of different types of EOSIO smart contract vulnerabilities.

In this work, we proposed EOSFuzzer, a general blackbox fuzzing framework to detect vulnerabilities within EOSIO smart contracts. EOSFuzzer includes a fuzzing input generator, a fuzzing executor, an instrumented Wasm VM, and a vulnerability detection engine. The fuzzing input generator can generate inputs based on the ABIs of the smart contracts and realize specific attacking scenarios with the agent smart contracts. The fuzzing executor can efficiently execute the generated inputs against the smart contract under fuzzing. The instrumented Wasm VM will collect opcode execution, API invocation and other execution information valuable for general vulnerability detection. Finally, the vulnerability detection engine implements the proposed test oracles to report vulnerabilities. Our experiment on 3963 smart contracts showed that EOSFuzzer is effective to detect vulnerabilities within EOSIO smart contracts with high accuracy. In particular, we have successfully mounted attacks on smart contracts without source code using EOSFuzzer to make a bet without spending any EOS in our case study.

[*] This research is supported in part by the National Key R&D Program of China under Grant 2019YFB2102400, NSFC (project no. 61772056),, the GRF of Research Grants Council (project no. 11214116, 11200015, and 11201114).
[#] EOSFuzzer is Open Sourced. https://github.com/gongbell/EOSFuzzer
[†] All correspondence should be addressed to Bo Jiang. E-Mail: gongbell@gmail.com.

The contributions of this work are three-fold. First, to the best of our knowledge, this work proposes the first fuzzing framework for detecting security vulnerabilities within EOSIO smart contracts. Second, the work proposes effective attacking scenarios and the corresponding test oracles that can trigger and detect typical vulnerabilities within EOSIO smart contracts with high accuracy. Third, we have systematically evaluated EOSFuzzer with a fuzzing experiment on real world EOSIO smart contracts, and EOSFuzzer has effectively identified more than 450 vulnerabilities within EOSIO smart contracts.

The organization of the remaining sections is as follows. In Section 2, we will present the basics of WebAssembly (Wasm), EOSIO platform and EOSIO smart contracts. Then in Section 3, we will present 3 typical vulnerabilities of EOSIO smart contracts. In section 4, we will present the design of our EOSFuzzer framework in detail. After that, we will perform a comprehensive experimental study to evaluate the effectiveness and efficiency of EOSFuzzer in terms of vulnerability detection in Section 5. Then we present case studies about successful attacks on EOSIO smart contracts with EOSFuzzer in Section 6. Finally, we present related works, acknowledgements, and conclusions in Section 7, 8, and 9.

## 2 Background

In this section, we provide background knowledge on WebAssembly, EOSIO platform, and the EOSIO smart contracts.

### 2.1 The WebAssembly

WebAssembly (Wasm for short) [32] is a binary instruction format for a stack-based virtual machine. Wasm is designed as a portable target for compilation of high-level languages like C/C++/Rust, enabling deployment on the web for client and server applications. Wasm is designed to be fast, safe, and ease of debugging. And it can be easily embedded in both web and non-web execution environment. The EOSIO platform also adopts Wasm VM for executing its smart contract.

### 2.2 EOSIO Platform and EOSIO Smart Contracts

EOSIO platform is an open source public blockchain platform that focuses on the scalability of transaction speed. The EOS is not only the token of EOSIO platform, but it also represents the stake hold by its owners. The EOSIO platform adopts the Delegated Proof of Stake (DPoS) protocol for making consensus, which is much more scalable than the Proof of Work (PoW) consensus protocol used by other platforms such as Bitcoin. The computing resources is also distributed according to the EOS tokens owned by the users. Moreover, EOSIO platform has set an upper limit on the execution time of each transaction. Exceeding the resource limit will lead to the rollback of the transaction with exceptions.

The features and characteristics of the EOSIO blockchain platform are designed to be flexible such that they can be modified to suit specific business requirement. Core blockchain features such as consensus, fee schedules, account creation and modification, token economics, etc., are implemented inside system smart contracts [30], which are deployed on the EOSIO blockchain platform. For example. The *eosio.token* system smart contract [24] defines the structures and actions allowing users to issue and transfer tokens on EOSIO based blockchains. The management of the EOS token is also performed by *eosio.token* system smart contract.

Smart contracts are programs running on the blockchain platform to help manage the assets of end users or to enforce the negotiation and execution of agreements among peers participating in the blockchain platform. The smart contracts [29] deployed on EOSIO platform contain two parts: the WebAssembly bytecode and the Application Binary Interface (ABI). The smart contract source code is compiled into Wasm bytecode for execution within the Wasm VM. And the ABIs describe the public interfaces of the smart contract to interact with.

Every EOSIO smart contract must provide an *apply* function as the entrance function to handle actions. The *apply* function will listen to all incoming actions and invoke the corresponding action handler functions accordingly. For example, the *transfer* function of a smart contract is usually used to handle *transfer* actions related to the contract [31].

The *apply* function uses the *receiver*, *code*, and *action* input parameters as filters to map the actions to the corresponding functions to handle [25]. The *receiver* is the account currently executing code. The *code* is the account that the action was originally sent to. And the *action* is the name of the action. It is important to understand the difference between code and receiver. To be specific, code is always the first receiver of the action, while receiver is the account currently executing the action. During the execution of a smart contract it may forward the action received to other account with the *require_recipient()* function. The *code* and *action* parameters of the *apply* function are also forwarded to the new contract. As a result, the *receiver* has changed to the newly notified contract, but the *code* stays the same.

## 3 Security Bugs in EOSIO Smart Contracts

In this section, we will briefly review the security bugs in EOSIO smart contracts.

### 3.1 Fake EOS Transfer

**Table 1. The EOSBet Contract with Fake EOS Transfer Vulnerability**

| | |
|---|---|
| 1 | … |
| 2 | *//The apply function* |
| 3 | **void** apply(**uint64_t** receiver, **uint64_t** code, **uint64_t** action){ |
| 4 | **auto** self = receiver; |
| 5 | if (code == self \|\| code == N(eosio.token) \|\| action == N(onerror)){ |
| 6 | TYPE thiscontract (self); |
| 7 | **switch**( action ) { |
| 8 | **EOSIO_API**(TYPE, MEMBERS) |
| 9 | }}} |

In the expected scenario, an EOSIO smart contract will only accept EOS transferred via the *eosio.token* system contract. If the contract under attack is vulnerable in that it does not check the *code* is *eosio.token* when the action is *transfer* within its *apply* function, an attacker may call the *transfer* function within the contract directly to fake an EOS transfer. As a result, the vulnerable contract may wrongly consider that the attacker has transferred EOS to it.

As shown in Table 1, within the *apply* function of the *EOSBet* contract, it only checks whether the code is the contract itself or eosio.token, but it does not check whether the action *transfer* is originally sent to the *eosio.token* system smart contract. As a result, an attacker contract may directly call the transfer function of the vulnerable contract to make a bet without spending any EOS.

**Table 2. Attacker Contract of the Fake EOS Transfer Vulnerability**

| 1 | **class** hackcontract : **public** eosio::contract { |
| --- | --- |
| 2 | … |
| 3 | **void** hack(account_name self, eosio::asset quantity, std::string memo) { |
| 4 | eosio::action( |
| 5 | eosio::permission_level{_self, N(active)}, |
| 6 | N(eosbet), |
| 7 | N(transfer), |
| 8 | std::make_tuple(self, N(eosbet), quantity, memo)).send(); |
| 9 | }}; |

As shown in Table 2, a hacker can use the attacker contract to exploit the fake EOS transfer vulnerability. The attack is simple: it directly performs an inline call to the transfer function of EOSBet.

**Table 3. The Fix of the Fake EOS Transfer Vulnerability**

| 1 | … |
| --- | --- |
| 2 | *//The apply function* |
| 3 | **void** apply(**uint64_t** receiver, **uint64_t** code, **uint64_t** action){ |
| 4 | **auto** self = receiver; |
| 5 | if( code == self || code == N(eosio.token) || action == N(onerror)){ |
| 6 | **if**( action == N(transfer) ){ |
| 7 | eosio_assert( code == N(eosio.token), "Must transfer EOS"); |
| 8 | } |
| 9 | … |

As shown in Table 3, The recommended way to fix the vulnerability is to add a check in the *apply* function to ensure *eosio.token* was the original receiver of the transfer action (line 6 and 7). In another word, when the action is *transfer*, the code must be *eosio.token*.

### 3.2 Forged Transfer Notification

When the transfer function of a contract is called, a smart contract may never check the destination (i.e., to) of the transfer. However, this is vulnerable since the smart contract may just receive a notification of a transfer to another contract rather than to itself.

During a normal EOS transfer, a receiver contract can choose to forward the *transfer* notification to other accounts through *require_recipient* (i.e. send a carbon copy of the *transfer* action). During a forged transfer notification attack [22], the attacker controls two accounts A and B. Then the attacker initializes the attack by transferring EOS from A to B through the system contract *eosio.token*. When the transfer is successful, both A and B will receive transfer notification. However, the contract deployed within account B can deliberately forwards the transfer notification to another contract (*eosbetcasino* in this example) with the *require_recipient* function in order to mislead it as shown in Table 4.

**Table 4. The Attacker Code to Send Forged Transfer Notification**

| 1 | class contractB : public eosio::contract { |
| --- | --- |
| 2 | public: |
| 3 | contractB(account_name self) : eosio::contract(self) {} |
| 4 | void transfer(uint64_t sender, uint64_t receiver) { |
| 5 | **require_recipient**( N(eosbetcasino)); |
| 6 | } |
| 7 | }; |

As shown in Table 5, the *eosbetcasino* is vulnerable in that it does not check whether the destination (i.e., *data.to*) of EOS transfer is itself within its transfer function. Then, it may wrongly consider that it has received the EOS. As a result, it may wrongly credit the attacker account A, who in fact has sent nothing to *eosbetcasino*.

**Table 5. The Smart Contract eosbetcasino**

| 1 | class eosbetcasino : public eosio::contract { |
| --- | --- |
| 2 | public: |
| 3 | void transfer(uint64_t sender, uint64_t receiver) { |
| 4 | auto data = unpack_action_data<st_transfer>(); |
| 5 | if (data.from == _self) *//no check for data.to* |
| 6 | return; |
| 7 | doSomething(); |
| 8 | } |
| 9 | }; |

As shown in Table 6, the fix of the forged transfer notification vulnerability is to simply check the destination of the transfer (line 5). If the destination of the EOS transfer is not the current contract, then the contract should directly ignore the transfer notification.

**Table 6. The Fix of the Forged Transfer Notification Vulnerability**

| 1 | class eosbetcasino : public eosio::contract { |
| --- | --- |
| 2 | public: |
| 3 | void transfer(uint64_t sender, uint64_t receiver) { |
| 4 | auto data = unpack_action_data<st_transfer>(); |
| 5 | if (data.from == _self || **data.to != _self**) |
| 6 | return; |
| 7 | doSomething(); |
| 8 | } |
| 9 | }; |

### 3.3 Block Information Dependency

Due to the lack of source of randomness, an EOSIO smart contract may rely on the block information such as *tapos_block_prefix* and *tapos_block_num* to generate random numbers, which may in turn determine the transfer of EOS or the winner of a lottery. However, the *tapos_block_prefix* and *tapos_block_num* are not reliable source of randomness, because they can be directly calculated from *ref_block_num*, which is the id of the last irreversible block by default.

A gambling contract may use deferred action to determine the winner of a lottery. In such scenario, the reference block is the block just before the block making the bet. Therefore, when a smart contract uses *tapos_block_prefix* and *tapos_block_num* directly for random generation, the random number generated can be predicted.

As shown in Table 7, the EOSDice smart contract [28] uses *tapos_block_prefix*, *tapos_block_num*, *current_time*, *account_name*, *game_id*, *pool_eos* for random number generation. Unfortunately, it turns out all these variables can be determined before taking the bet. The *account_name* (name of the contract), *game_id*, and *pool_eos* (balance of the current contract) are trivial to get. The *current_time* refers to the timestamp when determine the winner of the lottery, which is the sum of the timestamp of making the bet and the delay time of the deferred action. As a result, all variables used for random number generation can be computed before making the bet. Finally, the attackers successfully calculated the random number and won 2,545 EOS through the lottery, which was about $13,500.

**Table 7. EOSDice Contract with Block Info. Dependency Vulnerability**

| 1 | **uint8_t random(account_name name, uint64_t game_id){** |
|---|---|
| 2 | … |
| 3 | **auto mixd = tapos_block_prefix() * tapos_block_num() + name + game_id - current_time() + pool_eos.amount;** |
| 4 | **const char *mixedChar = reinterpret_cast<const char *>(&mixd);** |
| 5 | **checksum256 result;** |
| 6 | **sha256((char *)mixedChar, sizeof(mixedChar), &result);** |
| 7 | **uint64_t random_num = *(uint64_t *)(&result.hash[0]) + *(uint64_t *)(&result.hash[8]) + *(uint64_t *)(&result.hash[16]) + *(uint64_t *)(&result.hash[24]);** |
| 8 | **return** (uint8_t)(random_num % 100 + 1); |
| 9 | } |

## 4 The Design of EOSFuzzer

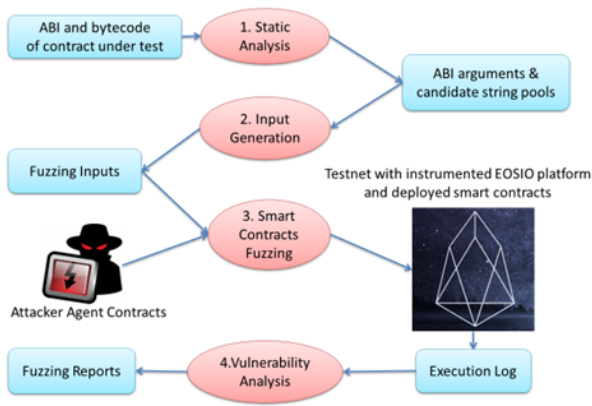

**Fig. 1 The Workflow of EOSFuzzer**

In this section, we will first present an overview of the EOSFuzzer. Then we will discuss each component of the EOSFuzzer in detail.

### 4.1 Overview of EOSFuzzer

As shown in Fig. 1, the general workflow of EOSFuzzer mainly consists four steps. First, EOSFuzzer performs static analysis on the ABI (application binary interface) and the bytecode of the smart contract under test. The static analysis step outputs the data types of the arguments for each ABI interface as well as a candidate string pool for string data type and memo parameter. Second, the EOSFuzzer performs input generation to generate fuzzing inputs for each ABI interfaces based on the static analysis result. Meanwhile, the EOSFuzzer will also use deployed attacker agent contracts to generate attacking scenarios for triggering specific vulnerabilities. Third, the EOSFuzzer performs fuzzing on the smart contract under analysis within the testnet through the cleos tool [23]. Note that the testnet is an instrumented EOSIO platform with smart contracts deployed on it. Through instrumentation, the execution information of the smart contract under analysis will be recorded into the execution log. Fourth, the EOSFuzzer will perform vulnerability analysis based on our proposed test oracles.

### 4.2 The Input Generator

The input generator is responsible for generating the inputs for the ABI interfaces of EOSIO smart contracts as well as generating the attacking scenarios with the attacker agents. The final inputs for fuzzing are the interleaved ABI invocations and the attack scenarios.

**Generating Inputs for Different Data Types**. The ABI interfaces for EOSIO smart contracts are in JSON format. Therefore, our tool parses the JSON file to extract the data type for each parameter. Then EOSFuzzer generates inputs for each data type and concatenates them to build the final input for invoking the ABI interfaces.

The primitive numeric data types include the 8-, 16-, 32-, and 64-bit integer and unsigned integer types as well as the 32- and 64-bit float-point types. For those data types, EOSFuzzer can randomly generate the maximum, or the minimum, or a random value within the input domain.

For the Boolean data type, the input generation algorithm will random return *true (1)* or *false (0)* values. For the string data type, our input generation algorithm will first build a candidate pool of strings with static analysis. More specifically, EOSFuzzer will iterate the data areas defining the constant data and collect all constant strings in the area as a pool of candidate strings. Then the EOSFuzzer will randomly return a string from the candidate pool for a string parameter during input generation.

For the *asset* data type, the algorithm will return random number of EOS tokens. For the symbol data type, the algorithm will only return the type EOS. In another word, we will mainly target at smart contract functions related to EOS token rather than other tokens.

The *name* data type represents the name of smart contracts. Currently, our tool will return the name of the smart contract under test as input for the parameter of *name* data type. Similarly, for the *public_key* data type, we will return the address of the current smart contract under test. This is because many of the ABIs are only

EOSFuzzer: Fuzzing EOSIO Smart Contracts for Vulnerability Detection

authorized to work on the current smart contract under test. The use of the *name* and *public_key* of other relevant smart contracts may further require the analysis of the relationships among smart contracts within a project, which we left as a future work.

For the *array* data type, our algorithm will iteratively generate inputs for its members of primitive data type. For *struct* data type, our algorithm will iteratively generate inputs for each of its members. If its member is of *struct* data type, the algorithm will also proceed recursively until primitive types are encountered.

For each smart contract, the input generation algorithm will generate one test suite containing a set of test cases. Each test case is a combination of ABI function invocations and attacking interactions (from the agent contracts) to the smart contract under fuzzing. The number of function invocations and attacking interactions within a test case is of random length and can be configured within EOSFuzzer.

**Generating Seeds for the Memo Parameter**. The memo argument is an argument used in the *transfer* function. Based on our analysis, we found that the sender and the receiver may use a memo argument of *string* type to confirm and guard a transfer action. To generate seeds for the memo parameter, our algorithm will perform static analysis on the smart contract under test to extract the strings used in the transfer function.

**Table 8. The Static Analysis to Extract Candidate Memo Strings**

| | |
|---|---|
| 1 | vector<string> getMemoStrings(vector wast, string param) { |
| 2 |   vector<string> memos |
| 3 |   set<string> offsets |
| 4 |   *for* func *in* all functions with signature like param |
| 5 |     *for* instruction contains ""i32.const"" { |
| 6 |       extract offset from instruction |
| 7 |       offsets.insert(offset) |
| 8 |     } |
| 9 |   *for* offset *in* offsets |
| 10 |     *foreach* instruction in data area contains offset { |
| 11 |       extract string from instruction |
| 12 |       memos.insert(string) |
| 13 |     } |
| 14 |   return memos |
| 15 | } |

As shown in Table 8, the function *getMemoStrings* extracts the candidate strings that may be used to compare with the *memo* argument within the *transfer* function. The *wast* parameter is the WebAssembly text format of the smart contract under fuzzing. EOSFuzzer uses the *wasm2wat* tool [33] to translate *wasm* bytecode to *wast* representation. The *param* is a string representing the data types of the return value and parameters of the transfer function (i.e., the string "(param i32 i64 i64 i32 i32)"). At line 4, the algorithm iterates all functions with signatures like the *transfer* function (i.e., having the same type of parameters and return value). Within each function, it iterates each instruction containing the use of constant values in the code (i.e., containing i32.const), and it extracts the offset of the corresponding constant variable in such functions (line 4 to 7). Finally, the algorithm iterates the data areas defining the constant data, locates the string corresponding to the offset (if it is of type *string*), and saves it in the vector memos as candidates for the memo parameter (line 9 to 12).

When there are functions with the same signature as the *transfer* function or when there are other constant strings used in the *transfer* function which are not related with *memo* parameter, the analysis may become imprecise. However, our analysis with the open source contracts showed that the *memos* vector only contains small number of irrelevant candidates, which is still effective for fuzzing.

**The Design of Attacker Agents for Faked EOS Transfer**. A safely written smart contract will check the *code* is eosio.token when the action is *transfer* within the apply function. Therefore, to trigger the fake EOS transfer vulnerability, we used an attacker contract called *fakeTransferAgent* to invoke the *transfer* function of the smart contract under fuzzing with *code* different from *eosio.token*. The invocation involves two scenarios and both are performed within the transfer function of the *fakeTransferAgent*. One is to invoke the transfer function with *code* equal to the receiver (i.e. the contract under fuzzing) while the other is to invoke the transfer function with *code* different from both the receiver and *eosio.token*. As a result, the agent contract will perform two possible attacks on the smart contract under fuzzing.

One attack is to perform an inline function call to the *transfer* function of the smart contract under fuzzing. In another word, our tool deliberately avoids using the *eosio.token* contract to perform the *transfer* action. In this attack, the *code* is equal to receiver (i.e., the smart contract under fuzzing).

The other kind of attack is to first use the *cleos* tool to call the *transfer* function of the *fakeTransferAgent* smart contract. Then, the *fakeTransferAgent* smart contract will forward the action to the smart contract under fuzzing through *require_recipient*. In this way, the smart contract under fuzzing will receive an transfer action with *code* equal to *fakeTransferAgent*, which is different from both *eosio.token* and itself (the receiver).

Finally, if the transfer function of smart contract under fuzzing is successfully triggered by the attack, it will be considered vulnerable.

**The Design of the Attacker Agents for Forged Transfer Notification**. To effectively trigger the *forged transfer notification* vulnerability, we designed a notifier agent contract to mount the attack. To initiate the attack, EOSFuzzer will first send some EOS to the notifier contract through the *eosio.token* smart contract. The notifier contract in turn will forward the transfer action to the smart contract under fuzzing through the *require_recipient()* function. Finally, the vulnerability detection module of EOSFuzzer will analyze the behavior of the smart contract under fuzzing on receiving the notification to determine whether it is vulnerable or not.

### 4.3 The Fuzzing Execution Engine

Having generated the inputs for fuzzing the smart contracts, the fuzzing execution engine is responsible for executing the inputs against the EOSIO smart contract. The fuzzing execution engine makes use of the *cleos* tool to interact with the smart contract. Within our design, the EOSFuzzer creates a pipe to communicate with the *cleos* process. It invokes the *push action* command of *cleos* tool to interact with the ABI interfaces of the smart contract under fuzzing or with the agent contracts.

For each smart contract, the input generation algorithm will generate a set of test cases. Each test case is a combination of ABI function invocations and attacking interactions (from the agent contracts) to the smart contract under fuzzing. When a test case finishes execution, EOSFuzzer will reset the testnet. On the one hand, resetting the testnet will lead to extra time cost. On the other hand, resetting the testnet will make the debugging easier since the execution trace is shorter. In this way, we want to make a tradeoff between fuzzing efficiency and debuggability after fuzzing.

When the execution of test cases ends, EOSFuzzer will start reading the instrumentation logs generated during execution and invoke the vulnerability detection module to analyze and report the vulnerability detection results.

### 4.4 Wasm VM Instrumentation

To support the extensible analysis of vulnerabilities, we must carefully instrument the Wasm VM to collect information about opcodes and operands invoked during execution. This includes those control-flow related opcode, the unary opcode, the binary opcode, and their operands. Whenever an opcode is executed, the instrumentation will generate a line recording the opcode, the operands, and the results (if there is any) within the opcode log file. Although not all information within the opcode log file are used for detecting the vulnerabilities studied in this work, the instrumentation log is design to be general for the analysis of new vulnerabilities in the future.

In particular, the *CallIndirect* opcode is instrumented because each invocation of the *execute_action* within the *apply* function will lead to the execution of *CallIndirect* opcode. Therefore, this opcode implies the invocation of an ABI function within the smart contract under test. Furthermore, we have also instrumented the Wasm API interface implementation code to collect information on whether the block information is queried during smart contract execution. To be specific, we have instrumented the *tapos_block_num()* and the *tapos_block_prefix()* functions.

Finally, we have also instrumented the *transfer* function of *eosio.token* system smart contract, which is used in EOSIO platform for EOS transfer. Instrumenting this function will help us understand whether EOS transfer actually happened during smart contract execution.

### 4.5 Test Oracles for Vulnerability Detection

In this section, we present the test oracles for vulnerability detection.

**Fake EOS Transfer**. To trigger the fake EOS transfer vulnerability, we used an attacker contract called *fakeTransferAgent* to perform two possible actions within its own transfer function. One is to directly perform an inline function call to the *transfer* function of the smart contract under fuzzing while the other is to forward a transfer action to the smart contract under fuzzing with *require_recipient*.

The test oracle to detect the fake EOS transfer vulnerability under the designed attacking scenarios is:

*CanReceiveEOS & TransferCalled*

The *CanReceiveEOS* test oracle is to check whether the smart contract can receive EOS rather than other tokens. During our implementation, we send some EOS to the smart contract under fuzzing through the eosio.token contract, and then we check whether the transfer function of the smart contract is invoked. Because at least two member functions will be called during the attacking scenario (i.e., the *transfer* function of *eosio.token* and the *transfer* function of the smart contract under fuzzing), EOSFuzzer will check whether the *CallIndirect* opcode has been executed at least 2 times within its implementation accordingly.

The *TransferCalled* test oracle is to check whether the *transfer* function of the smart contract under fuzzing can be successfully triggered by the *fakeTransferAgent*. Similar to *CanReceiveEOS*, EOSFuzzer checks whether the *CallIndirect* opcode has been executed for at least 2 times (i.e., one for the *transfer* function of *fakeTransferAgent* and the other for the *transfer* function of the smart contract under fuzzing). Since the attacking scenario is well controlled by our agent contract, we can be sure that the transfer function is the only ABI function invoked in the smart contract under fuzzing. Therefore, the appearance of the second *CallIndirect* opcode corresponds to the invocation of the *transfer* function rather than other ABI functions within the smart contract under fuzzing.

**Forged Transfer Notification**. As presented in previous sections, to detect *forged transfer notification* vulnerability, the EOSFuzzer makes use of a notifier agent to send a forged notification to the smart contract under fuzzing.

The test oracles to detect forged transfer notification is as follows: *TransferCalled & !CheckRecipient*

The *TransferCalled* sub-oracle is to check whether the *transfer* function of the smart contract under fuzzing is invoked during the attacking scenario. Within the call chain of our attack scenarios, if the *transfer* function of the smart contract is called, at least three ABI functions must be invoked. These corresponds to the *transfer* functions of *eosio.token* contract, the *notifier* contract, and the smart contract under fuzzing. Each invocation of a member function will lead to the execution of a *CallIndirect* opcode. Therefore, when realizing the oracle *TransferCalled*, EOSFuzzer checks whether there are at least 3 *CallIndirect* opcode executed during our attack. At the same time, EOSFuzzer also records the line number of the third *CallIndirect* opcode within the opcode log file for ease of further analysis, which corresponds to the starting position of the *transfer* function within the smart contract under fuzzing.

The *!CheckRecipient* condition is to check the smart contract under fuzzing has not checked whether it is the real recipient of the EOS transfer in **all** attacks by the notifier agent. To realize the sub-oracle *!CheckRecipient*, EOSFuzzer first checks whether there is any execution of comparison opcode (i.e., Eq or Ne) between the notifier contract (the actual receiver of EOS) and the smart contract under fuzzing in one round of attack by the notifier agent. The comparison starts from the starting position of the *transfer* function of the smart contract under fuzzing recorded in the opcode log file during the previous step. It should be noted that directly checking the instruction on comparing the *to* and *_self* parameters of the transfer function may be imprecise because EOSIO does not enforce the consistency between the ABI interface and the actual implementation.

During our fuzzing process for a smart contract, EOSFuzzer will initiate the attack from the notifier agent many times, the *!CheckRecipient* condition is satisfied only when the check of the EOS transfer recipient is not performed in **all** attacks from the notifier agent. Since a Fuzzer may not traverse each path to the code for recipient checking, EOSFuzzer may produce false positives when reporting Forged Transfer Notification vulnerability. However, when the scale of the fuzzing is large, the probability of generating false positives will become very low, which is also confirmed by our experiment evaluation.

**Block Information Dependency**. The test oracle to detect block information dependency is as follows:

*BlockInfoRead & EOSTransferCalled*

The *BlockInfoRead* test oracle checks whether there is any invocation of APIs reading block information. During the implementation, this is realized through instrumenting the *tapos_block_num* and *tapos_block_prefix* API. The *EOSTransferCalled* oracle checks whether the *transfer* function of the *eosio.token* smart contract is called. The EOSFuzzer realizes this by simply instrumenting the transfer function of *eosio.token* system smart contract and re-deploying it within EOSIO.

### 4.5 Discussions

In this section, we discuss the possible extensions to EOSFuzzer in two aspects.

First, EOSFuzzer is designed as a blackbox fuzzer, but it can be also extended to be a grey-box fuzzer. The key to the extension is to collect code coverage information during fuzzing process. Currently, EOSFuzzer instruments all kinds of bytecode instructions within WASM VM. We can extend EOSFuzzer by collecting branch coverage information within VM instrumentation: whenever a branch instruction is encountered, we can log the location of the two branches as well as the coverage of new branch. The incremental branch coverage information can be used to guide the mutation-based grey-box fuzzing process. New mutation and selection strategies can be proposed to further improve the effectiveness of the fuzzer.

Second, EOSFuzzer can also be extended to support the detection of new vulnerabilities. For example, the rollback vulnerability [10] is a vulnerability within the smart contracts of gambling DApp. In essence, the vulnerability lies in the *reveal* function of the gambling smart contract where all the actions are located in a single transaction. As a result, an attacker can revert the whole transaction to restore its balance whenever it loses the gambling game. We can extend EOSFuzzer to use another attacker agent to realize the above attacking scenarios. The transfer of EOS to the contract under fuzzing, the invocation of the *reveal* function, the checking of the balance of the attacker agent, and the invocation of the revert operation can all be easily supported by EOSFuzzer framework through simple extension.

## 5 Experiment and Results Analysis

In this section, we present our fuzzing experiment to evaluate the vulnerability detection effectiveness of the EOSFuzzer.

### 5.1 Subject Programs

The subject programs used in our experimental study include 82 open source EOSIO smart contracts and 3881 smart contracts without source code. The 82 open source smart contracts are mainly used to manually verify the effectiveness of our tool. However, collecting large number of EOSIO smart contracts with source code is difficult because most EOSIO DApps are not open-source. The 3881 smart contracts without source code contain Wasm bytecode and ABI file only, which are used to evaluate our tool at scale. For the smart contracts with source code, we first compiled them into bytecode and then deployed them in our testnet for fuzzing. For the smart contracts with Wasm bytecode only, we directly deployed them within our testnet.

### 5.2 Experiment Setup

The experiment was performed within a docker. The host for the docker was a desktop equipped with 8 cores of Intel® Core™ i7-6700 CPU @ 3.4GHz and 16GB of memory. The host was running Ubuntu 16.04.1 and Docker version 18.06.1. Furthermore, the docker was running Ubuntu 18.04.4. We ran the instrumented EOSIO client (node) and the EOSFuzzer within the Ubuntu OS in the docker. The EOSIO client used is the nodeos version 1.5.2. And the EOSFuzzer talks to the EOSIO node through the cleos tool.

Before performing the fuzzing experiment, we must properly configure the EOSIO testnet. First, we started the *keosd* wallet to provide key manager service daemon for storing private keys and signing digital messages. Second, we ran the instrumented nodeos daemon to start the single node testnet. Third, we deployed the instrumented eosio.token system smart contract and the attacker agent contracts. Fourth, we created 2 EOSIO accounts to cooperate with the attacker agent contracts for performing EOS transfer during attack. Finally, we deploy the smart contracts under fuzzing on the EOSIO testnet. The deployed smart contracts include 3881 smart contracts without source code and 82 smart contracts with source code.

When the EOSIO testnet had been configured, we further configured EOSFuzzer to perform around 1000 interleaved ABI function invocations and attacks from agent contracts in total for each smart contract. Finally, we started to perform fuzzing with the EOSFuzzer against the deployed smart contracts.

### 5.3 Experiments and Results Analysis

In this section, we first present the results of EOSFuzzer on contracts with and without source code. Then we further compare the EOSFuzzer and EVulHunter tool [17] in terms of vulnerability detection effectiveness. The EVulHunter is an EOSIO smart contract vulnerability detection tool based on static analysis. Finally, we present the efficiency of EOSFuzzer in terms of vulnerability detection.

**EOSFuzzer Results on Contracts with Source Code**. The vulnerability detection result of EOSFuzzer is shown in Table 9. The column *Total* represents the total number of smart contracts under analysis. The column *Reported*, *FP*, and *FN* represents the number of vulnerabilities, false positive cases and false negative cases reported for the corresponding vulnerability type, respectively. We manually checked the source code of the 82 smart contracts to identify the false positive or false negative cases.

**Table 9. Results of EOSFuzzer on Smart Contracts with Source Code**

| Vulnerability | Total | EOSFuzzer | | |
|---|---|---|---|---|
| | | Reported | FP | FN |
| Block Info. Dependency | 82 | 2 | 0 | 1 |
| Forged Transfer Notification | 82 | 4 | 0 | 0 |
| Fake EOS Transfer | 82 | 2 | 1 | 0 |

Among the 82 smart contracts, the EOSFuzzer has detected 2 block dependency vulnerabilities. Both of them are confirmed with our manual check. There is one false negative case by EOSFuzzer for the block dependency vulnerability.

The smart contract *lottery10* is the false negative case of EOSFuzzer. As shown in Table 10, the *transferact* function is designated to handle the transfer of EOS for making a bet. However, the contract only accepts EOS transfer with some pre-determined amount (i.e., 1000, 10000, 100000…). Moreover, the contract further requires the 10$^{th}$ invocation of the transfer function with the pre-determined amount can actually make the bet, which makes the EOSFuzzer hard to trigger without a large-scale fuzzing effort. In general, performing a very large-scale fuzzing campaign against a smart contract or using feedback mechanisms for fuzzing may help expose the vulnerability, which is left as a future work.

**Table 10. False Negative Case of Block Info. Dependency by EOSFuzzer**

| 1 | class lottery10 : public eosio::contract { |
|---|---|
| 2 | … |
| 3 | void transferact(uint64_t receiver, uint64_t code) { |
| 4 | //function to handle transfer |
| 5 | … |
| 6 | int64_t amount = data.quantity.amount; |
| 7 | if (amount == 1000 \|\| amount == 10000 \|\| amount == 100000 \|\| amount == 1000000 \|\| amount == 10000000) { |
| 8 | //accept the bet |
| 9 | … |
| 10 | }; |

For the Forged Transfer Notification vulnerability, the EOSFuzzer has identified 4 smart contracts. After manual checking all the 82 smart contracts, we confirmed that EOSFuzzer reports neither false positives nor false negatives.

For the Fake EOS Transfer Vulnerability, EOSFuzzer identified 2 vulnerabilities out of the 82 smart contracts. After manual checking, we confirmed that EOSFuzzer reports no false negatives. But one of the identified vulnerable contracts (called vigor) is a false positive case.

**Table 11. A False Positive Case of Fake EOS Transfer by EOSFuzzer**

| 1 | CONTRACT vigor: public eosio::contract { |
|---|---|
| 2 | … |
| 3 | [[noreturn]] void apply(uint64_t receiver, uint64_t code, uint64_t action) { |
| 4 | if((code==name("eosio.token").value \|\| |
| 5 | code==name("vig111111111").value \|\| |
| 6 | code==name("dummytokens1").value) && |
| 7 | action==name("transfer").value) { |
| 8 | eosio::execute_action(name(receiver),name(code),&vigor::assetin); |
| 9 | } |
| 10 | if (code == receiver) { |
| 11 | switch (action) { //typical action mapping workflow |
| 12 | EOSIO_DISPATCH_HELPER(vigor, (create) (transfer)…) |
| 13 | } |
| 14 | } |
| 15 | } |

As show in Table 11, the smart contract *vigor* is a false positive case of Fake EOS Transfer reported by EOSFuzzer. In typical workflow, the *apply* function will use the *transfer* function to handle the transfer of EOS token, which is also the assumption our test oracle implementation based on. However, within the vigor *contract*, the *assetin* function (rather than the *transfer* function) is used to handle the transfer of EOS (line 4 to 8). Therefore, the sub-oracle can *CanReceiveEOS* can be satisfied (i.e., *CallIndirect* opcode has been executed for at least 2 times). Furthermore, when the attacker agent performs an inline invocation of the transfer function on *vigor*, the *transfer* function can still be called through the default dispatcher macro (line 10 to 12). So, the sub-oracle *TransferCalled* can also be satisfied (since the *CallIndirect* opcode has been executed for at least 2 times). Since both sub-oracles are satisfied, our *EOSFuzzer* wrongly reports the contract as a *Fake EOS Transfer* vulnerability. The reason for the false positive is due to the fact that *vigor* uses a different function called *assetin* rather than the function *transfer* to handle EOS transfer, which are hard to know without analyzing the source code. However, we consider the use of functions other than *transfer* to handle EOS transfer as unusual for EOSIO smart contracts.

**Comparison of EOSFuzzer and EVulHunter**. In this section, we further compare EOSFuzzer with the EVulHunter tool for vulnerability detection as shown in Table 12. Since the EVulHunter tool does not the support the detection of Block Information Dependency vulnerability, we compare EOSFuzzer and EVulHunter in terms of the *Forged Transfer Notification* and *Fake EOS Transfer* vulnerability only.

**Table 12. Comparison of EOSFuzzer and EVulHunter**

| Vulnerability | Total | EOSFuzzer | | | EVulHunter | | |
|---|---|---|---|---|---|---|---|
| | | Reported | FP | FN | Reported | FP | FN |
| Forged Transfer Notification | 82 | 4 | 0 | 0 | 12 | 10 | 2 |
| Fake EOS Transfer | 82 | 2 | 1 | 0 | 9 | 8 | 0 |

Among the 82 smart contracts, the EVulHunter successfully analyzed 74 smart contracts. It fails to generate output for the other 8 contracts. Within the 74 smart contracts, The EVulHunter has reported 12 smart contracts with Forged Transfer Notification vulnerability. However, after manual check, we found 10 of them are false positives. Furthermore, EVulHunter missed 2 vulnerable smart contracts. Therefore, the EOSFuzzer tool outperforms the EVulHunter when detecting Forged Transfer Notification vulnerability.

As shown in Table 13, the smart contract *salescon* is a false positive case for forged transfer notification vulnerability by EVulHunter. It uses the assertion statements (i.e., eosio_assert) to perform conditional checks. However, the EVulHunter tool seems only works well with conditional checks using *if* statement, leading to false positives when assertion is used. During our manual confirmation, when we change the *assertion* statement to *if* statement in the contract, the EVulHunter tool will not generate the false positive any more.

**Table 13. A False Positive Case for Forged Transfer Notif. by EVulHunter**

| 1 | class [[eosio::contract]] salescon : public eosio::contract { |
| 2 | … |
| 3 | void salescon::transfer(name from, name to, asset quantity, string memo) |
| 4 | {… |
| 5 | eosio_assert(from == buyer, "Transfer must come from buyer"); |
| 6 | eosio_assert(to == _self, "Contract was not the recipient"); |
| 7 | eosio_assert(quantity.symbol.is_valid(), "Invalid quantity"); |
| 8 | … |

One false negative case for forged transfer notification vulnerability by the EVulHunter tool is the smart contract *vigor*. During its analysis, the EVulHunter tool fails to find any path leading to the indirect call. However, the EOSFuzzer tool did find a path leading to the indirect call during the fuzzing process. Therefore, the EVulHunter seems not precise enough during its static analysis process.

For the Fake EOS Transfer Vulnerability, EOSFuzzer identified 2 vulnerabilities while the EVulHunter tool reported 9 vulnerabilities. However, after manual check, we found 8 of the 9 vulnerabilities reported by EVulHunter are actually false positives. In contrast, only 1 of the 2 vulnerabilities reported by EOSFuzzer is false positive. Therefore, for Fake EOS Transfer Vulnerability, EOSFuzzer incurs much lower false positives than EVulHunter.

As shown in Table 14, the *token* smart contract is reported by EVulHunter as *Fake EOS Transfer* vulnerability. However, within the token contract, it uses the EOSIO_ABI macro for default action mapping. The EOSIO_ABI can only handle scenarios where *code* is equal to *self*. Therefore, the *token* contract cannot handle EOS transfer from the *eosio.token* contract. It is only a contract handling tokens other than EOS. Therefore, the *token* contract is a false positive case of EVulHunter.

**Table 14. A False Positive Case for Fake EOS Transfer by EVulHunter**

| 1 | class token: public contract { |
| 2 | … |
| 3 | }; |
| 4 | EOSIO_ABI( eosio::token, (create)(issue)(transfer) ) |
| 5 | *//macro for default action mapping* |

**EOSFuzzer Results on Contract without Source Code**. We have also performed fuzzing on 3881 EOSIO smart contract with bytecode and ABI only. As shown in Table 15, the EOSFuzzer has identified 3 Block Information Dependency vulnerabilities, 183 Forged Transfer Notification vulnerabilities, and 265 Fake EOS Transfer vulnerabilities. Among the 3881 smart contracts, the percentage of Block Information dependency, Forged Transfer Notification, and Fake EOS Transfer are 0.07%, 4.72%, and 6.83%, respectively. Without source code, it is hard to manually verify those reported smart contracts to identify the precise number of false positives and false negatives. But the results give us an estimation of the number of vulnerable smart contracts in the wild.

Considering the potential large number of Forged Transfer Notification and Fake EOS Transfer vulnerability, the developers are recommended to perform security check and harden their contract before releasing their EOSIO smart contract to public.

**Table 15. Results of EOSFuzzer on Smart Contracts without Source Code**

| Vulnerability | Total | Vulnerabilities Detected | Percentage |
|---|---|---|---|
| Block Info. Dependency | 3881 | 3 | 0.07% |
| Forged Transfer Notification | 3881 | 183 | 4.72% |
| Fake EOS Transfer | 3881 | 265 | 6.83% |

**Efficiency of EOSFuzzer**. When fuzzing each contract, we configured EOSFuzzer to perform around 1000 interleaved ABI function invocations and attacks from agent contracts in total for each smart contract. For each contract, the fuzzing experiment takes around 90 seconds on average, which is small based on the hardware configurations of our experiment. Furthermore, the average invocations per second is around 11, which is fast enough for comprehensive fuzzing. This is partly due to the high transaction speed of EOS based on the DPOS consensus protocol. Therefore, EOSFuzzer is also efficient for EOSIO smart contract vulnerability detection.

## 6 Mounting Attacks on Vulnerable Contracts

We have also successfully mounted an attack on a smart contract called *diamond1* to earn EOS within our testnet. Since *diamond1* provides no source code, we performed fuzzing on its bytecode based on its ABI interfaces. In addition to the test oracles defined within EOSFuzzer, we also checked whether the EOS balance of the *diamond1* is reduced after each ABI invocation (Initialized to be 1000 EOS before fuzzing). In another word, we wanted to ensure that the asset of the contract under fuzzing was lost through the attack.

Within 7 ABI invocations, the EOSFuzzer not only successfully triggered the *Forge Transfer Notification* vulnerability, but it also successfully made a bet and earned a reward (26.46 EOS) without using a single EOS. The detailed attacking process is as follows:

During the first ABI invocation, the EOSFuzzer performed an EOS transfer to the *diamond1* smart contract and checked whether its *transfer* function was invoked. This was to check whether *diamond1* could receive EOS. During the following 3 ABI invocations, the EOSFuzzer performed the *Forged Transfer Notification* attacks by transferring EOS to the notifier agent contract from a sender account. The notifier agent contract will send the notification to the *diamond1* smart contract through the *require_recipient* function. For all 3 invocations, the *Forged Transfer Notification* were all triggered, and EOSFuzzer successfully made a bet without spending any EOS. During the 5$^{th}$ invocation, the EOSFuzzer performed the *Fake EOS Transfer attack* with the agent contract to perform an inline invocation to the *diamond1* smart contract. However, no *Fake EOS Transfer* vulnerability was triggered. During the 6$^{th}$ invocation, EOSFuzzer invoked the *deposit* ABI function of *diamond1*. However, its balance was not changed. Finally, during the 7$^{th}$ ABI invocation, EOSFuzzer invoked the *endlottery* ABI function of the *diamond1* smart contract, which ended the betting process and announced the betting results. Note that the *endlottery* ABI function can also be triggered from the website of the gambling game.

At the end of the last invocation, EOSFuzzer found that the balance of *diamond1* smart contract reduced to 973.54 EOS while the balance of the sender account increased from 1000 to 1026.46. The logs further confirmed the success of the attack: the *diamond1* smart contract considered the sender account had successfully made a bet during the *Forged Transfer Notification* attack, it announced the *sender* account as the winner, and it had sent 26.46 EOS to the sender account as the reward. However, the sender account only sent the notifier agent contract some EOS for conspiracy, it earned 26.46 EOS without any stake.

## 7 Related Work

In this section, we present closely related work on smart contract vulnerability detection and fuzzing techniques.

### 7.1 Smart Contract Vulnerability Detection

Parizi et al. [15] carry out an extensive experimental assessment of current static smart contracts security testing tools for the Ethereum smart contracts. Tsankov et al. proposed the Securify tool for smart contract vulnerability detection. The Securify tool [18] is a scalable security analyzer for Ethereum smart contracts and it can prove contract behaviors as safe/unsafe with respect to a given property. Abdellatif and Brousmiche [1] used a formal modeling approach to verify a smart contract behavior in its execution environment. They further analyzed the security of contracts with a statistical model checking approach.

Luu et al [12] designed Oyente, a symbolic verification tool for Ethereum smart contract. Oyente builds the control-flow graph of smart contracts and then performs symbolic execution on the control flow graph while checking whether there exist any vulnerable patterns. Nikolic et al. [14] designed MAIAN, a symbolic execution tool for reasoning about tracing properties to detect vulnerable Ethereum smart contracts. It specified three typical smart contracts vulnerabilities based on trace properties. The MAIAN can efficiently detect the greedy, the prodigal and the suicidal contracts through symbolic execution. Hirai [11] used Isabelle/HOL tool to verify the smart contract called Deed, which is part of the Ethereum Name Service implementation. Specifically, the work verifies the oracle that only the owner of Deed could decrease its balance. Furthermore, they also found the EVM implementation is poorly tested during the verification process.

Chen et al. [5] proposed the TokenScope tool, which automatically checks whether the behaviors of the token contracts are consistent with the ERC–20 standards. Nguyen et al. proposed the sFuzz tool [7], which combined the strategy in the AFL fuzzer and an efficient lightweight multi-objective adaptive strategy targeting those hard-to-cover branches. The evaluation shows the sFuzz tool is efficient and effective to achieve high code coverage and to detect vulnerabilities. Wang et al. [19] proposed the VULTRON tool, which can precisely detect irregular transactions in smart contracts due to various types of adversarial exploits. It provided a general way to solve the test oracle problem for smart contract vulnerability detection.

The techniques reviewed above are in general effective to detect the vulnerabilities or bugs within Ethereum smart contracts. However, they are not specifically designed to support the detection of vulnerabilities within EOSIO smart contracts.

The EVulHunter [17] is a static analysis tool to detect vulnerabilities within EOSIO smart contracts. However, it fails to generate results for a number of EOSIO smart contracts and it generates many false negatives and false positives for vulnerability reporting. EOSAFE [10] is another static analysis framework to detect vulnerabilities within EOSIO smart contracts based on symbolic execution. We believe dynamic fuzzing and static analysis are two complementary techniques for EOSIO smart contract vulnerability detection.

### 7.2 Fuzzing Techniques for Vulnerability Detection

There are many works on fuzzing techniques for vulnerability detection.

GodeFroid [9] et al. proposed to enhance whitebox fuzzing of complex structured-input applications with a grammar-based specification of their valid inputs. Their test data generation algorithm combines symbolic execution and constraint solving to improve the fuzzing process. Wang et al. proposed TaintScope [20], an automatic fuzzing system using dynamic taint analysis and symbolic execution techniques. TaintScope can identify checksum fields in input instances, locate checksum-based integrity checks by using branch profiling techniques, and bypass such checks via control flow alteration.

Dai et al. [6] proposed the configuration fuzzing technique, which randomly modifies the configuration of the running application at certain execution points to check for vulnerabilities. Ganesh realized the BuzzFuzz tool [8], which uses dynamic taint tracing to automatically locate regions of original seed input files that influence values used at key program attack points. The BuzzFuzz tool then automatically generates new fuzzed test input files by fuzzing these identified regions of the original seed input files.

Chen et al. [3] proposed a mutation-based fuzzer called Angora, which proposed to use scalable byte-level taint tracking, context-sensitive branch count, search based on gradient descent, and input length exploration techniques to solve path constraints efficiently. Lyu et al. [13] proposed a novel mutation scheduling scheme, which made mutation-based fuzzer to discover vulnerabilities more efficiently. Chen et al. [4] proposed a globally asynchronous and locally synchronous (GALS) seed synchronization mechanism to seamlessly ensemble base fuzzer for better performance.

You et al. [21] proposed the ProFuzzer, which automatically recovered and understood input fields of critical importance to vulnerability discovery during a fuzzing process. The fuzzer can intelligently adapt the mutation strategy to enhance the chance of hitting zero-day targets.

In general, different from EOSFuzzer, the fuzzing techniques reviewed above are not specifically designed for detecting vulnerabilities within EOSIO smart contracts.

There are also interesting works on stateful fuzzing techniques. The AFLnet takes a mutational approach and uses state-feedback to guide the fuzzing process of network protocols [16]. The REST-ler [2] is a stateful automatic intelligent REST API security-testing tool. REST-ler generates tests intelligently by inferring dependencies among request types declared in the Swagger specification. It also analyzes dynamic feedback from responses observed during prior test executions in order to generate new tests. Currently, the fuzzing performed by EOSFuzzer is stateless. We may extend EOSFuzzer to perform stateful fuzzing with the support of data-flow analysis in the future work.

## 9 Conclusions and Future Work

The vulnerabilities within the EOSIO smart contracts have resulted in significant loss for its users. Therefore, effective tools for detecting the vulnerabilities within EOSIO smart contract are needed. In this work, we present EOSFuzzer, a black-box fuzzing tool to automatically detect vulnerabilities within EOSIO smart contracts. Within EOSFuzzer, we have proposed effective attacking scenarios as well as invocations to ABI interfaces to perform comprehensive fuzzing on EOSIO smart contracts. Furthermore, we have also defined and realized test oracles to detect three typical vulnerabilities for EOSIO smart contracts. Our fuzzing experiment showed that EOSFuzzer is effective and efficient to detect vulnerabilities within EOSIO smart contracts for practical use. In particular, EOSFuzzer has successfully mounted attacks on EOSIO smart contract without source code to make bets without spending any EOS.

For future work, we would like to extend the EOSFuzzer to support the detection of new types of vulnerabilities in EOSIO smart contracts. We also plan to improve the search strategies of EOSFuzzer during fuzzing to improve its vulnerability detection effectiveness for complex smart contract